\documentstyle[12pt]{article}
\begin{document}
\renewcommand{\thesection}{\normalsize\bf\arabic{section}}

\begin{center}
~\\~\\~\\~\\~\\

\large
An Analytical and Numerical Study of Optimal Channel Networks

 ~\\~\\ ~\\

\normalsize
F. Colaiori $^{\circ}$, A. Flammini$^{\circ}$, A. Maritan $^{\circ}$
$^{\dagger}$ and Jayanth R. Banavar$^{\ddagger}$

 ~\\ ~\\  ~\\

$^{\circ}$
{\it Istituto Nazionale di Fisica della Materia (INFM),}

{\it International School for Advanced Studies (ISAS),}

{\it via Beirut 2-4, 34014 Trieste, ITALY}

$^{\dagger}$
{\it Istituto Nazionale di Fisica Nucleare (INFN),}

{\it and}

$^{\ddagger}$
{\it Department of Physics and Center for Material Physics, The Pennsylvania
State University 104, Davey Laboratory, University Park, PA 16802, USA.}

 ~\\~\\~\\~\\

\end{center}

\newpage

~\\~\\~\\

\begin{abstract}
We analyze the Optimal Channel Network model for river networks using both  
analytical and numerical approaches. This is a lattice model in which a 
functional describing the dissipated energy is introduced and minimized in 
order to find the optimal configurations. The fractal character of river 
networks is reflected in the power law behaviour of various quantities 
characterising the morphology of the basin. In the context of a finite size 
scaling Ansatz,  the exponents describing the power law behaviour are 
calculated exactly and show mean field behaviour, except for  two limiting 
values of a parameter 
characterizing the dissipated energy, for which the system belongs to 
different universality classes. Two modified versions of the model, 
incorporating quenched disorder are considered: the first simulates 
heterogeneities in the local properties of the soil, the second considers 
the effects of a non-uniform rainfall. In the region of mean field behaviour, 
the model is shown to be robust to both kinds of perturbations. In 
the two limiting cases the random rainfall is still irrelevant, whereas 
the heterogeneity in the soil properties leads to new universality classes. 
Results of a numerical 
analysis of the model are reported that confirm and complement
the theoretical analysis of the global minimum.  
The statistics of the local minima are found 
to more strongly resemble observational data on real rivers.
\end{abstract}

\newpage

\section{\normalsize\bf Introduction}

Experimental observations on  river networks have shown clear evidence of 
their fractal character. Data from many basins with different geological 
features have been analyzed, and have shown power law behaviour of the
probability distributions of several quantities describing the morphology of 
the river basin \cite{a,d,n,p} over a wide range of scales.

\noindent 
Several statistical models have been proposed \cite{ai,f,ba,bb,bc,AA} but a 
complete theoretical understanding is still lacking. Recently a simple 
lattice model derived from an energy minimization principle has been 
proposed \cite{ac,ad,BB} that in spite of its simplicity, seems to reproduce 
many features of natural river networks.

Numerical investigations of the model have been performed \cite{f,h,ag,ah}
with different geometrical constraints on the form of the basin. Furthermore, 
the model has been analyzed within the framework of a finite size scaling 
Ansatz \cite{i}. In the present paper  the so called {\em Optimal Channel 
Network } model \cite{ac,ad} is studied analytically and exact results are 
obtained. 

\noindent
In addition, generalized models, taking into account the presence of quenched 
disorder are considered. Randomness is introduced in two different ways: one  
modelling the inhomogeneity of the soil and the other non-uniformity in the 
rainfall. Analytical results are given in these two cases, too.

\noindent
Results of numerical simulation in the last part of the paper, confirm and 
complement the analytic predictions. 

In Section 2 we describe the lattice model and derive the scaling laws. The 
relationship between exponents are also derived. The exponents characterizing 
the power law distributions of drained areas and mainstream lengths are 
expressed in terms of a single independent exponent: the wandering exponent 
in the self-affine case and the fractal dimension for the self-similar basin.
The section ends with the definition of the Optimal Channel Network model and 
with a short discussion of the underlying  minimization principle.

\noindent
Section 3 is entirely devoted to a analytical study of the model on a simple 
fractal, the Peano basin. The solution is given exactly and is used to give 
bounds in the subsequent Section. The distributions of areas and lengths are 
evaluated and shown to exhibit the  form predicted by the finite size scaling 
Ansatz. 

\noindent
In Section 4, analytical results in the absence of disorder are derived. The 
model is shown to exhibit three distinct universality classes for different 
values of a parameter characterizing the dissipated energy. 

\noindent
Heterogeneities in the soil properties and random rainfall are considered in 
the  generalized models of Section 5. Analytical results for these cases are 
also  deduced. 

\noindent
Numerical results pertaining to the search for the global minimum of the 
dissipated energy with a simulated annealing algorithm are given in Section 6 
and numerical results for the statistics of the local minima are given in 
section 7. 

\section{\normalsize\bf Definitions and Derivation of Scaling Laws}

A river basin is described by a scalar field of elevations. Drainage 
directions are identified by steepest descent, i.e. by the largest local 
decrease of the elevation field.
The presence of lakes  has not been considered, i.e. from each point 
the water can flow to one of the nearest neighbours. This corresponds 
to have all lakes saturated.  

\noindent
Within this context, a river network can be represented by an oriented 
spanning-tree over a two dimensional lattice of arbitrary size and shape, 
in which oriented links (one coming out from each site of the lattice) 
correspond to drainage directions.

\noindent
We will consider spanning trees rooted in a corner of  a $L \times L$ square 
lattice (the root will correspond to the outlet). Site $i$ is upstream 
respect to site $j$ if there is an oriented path from $i$ to $j$. To each 
site $i$ of the lattice, we associate a local injection of mass $r_{i}$ (the 
average annual rainfall in the site $i$). The flow $A_{i}$, also referred to 
as the accumulated area can thus be defined as the sum of the  injections
over all the points upstream of site $i$ (site $i$ included). 

\noindent
The variables $A_{i}$ are thus related by 

\begin{equation}
A_{i}= \sum_{j } w_{i,j} A_{j} + r_{i} \, ,
\label{aree}
\end{equation}

where $w_{i,j}$ is $1$ if site $j$ is upstream with respect to site $i$ and 
is a nearest neighbour of it and $0$ otherwise. The local injection $r_{i}$ 
is commonly assumed to be homogeneous and identically equal to $1$.

\noindent
In natural basins these areas can be investigated through an experimental 
analysis of digital elevation maps (DEM's) \cite{d}. See figure 1a for 
an example. 
\noindent

The upstream length relative to a site is defined as the length of the stream 
obtained starting from the site and repeatedly moving in the upstream 
direction towards the nearest-neighbour with biggest area $A$ (the one 
leading to the outlet is excluded, since it is a downstream site), until a 
source is reached i.e. a site with no incoming links (see Fig. 1b). If two or 
more equal areas are encountered, one is randomly selected.

\noindent
For a given tree, one may consider the probability distribution of the 
following quantities: for a lattice of given linear size $L$ we will call 
$p(a,L)$ the probability density of accumulated areas $a$ and $\pi(l,L)$ the 
probability density of the upstream lengths $l$. These represent  the fraction 
of sites with area $a$ or stream length $l$ respectively. We will also 
consider the integrated probability distributions $P(a,L)$, the probability 
to find an accumulated area bigger then $a$ and $\Pi(l,L)$, i.e. the 
probability to have a site with an upstream length bigger than $l$. 

\noindent
Both these probability distributions, here defined in the simple case of
the lattice model, were originally introduced to describe real rivers
and experimentally found to scale as power laws leading to the formulation 
of a finite size scaling ansatz\cite{i}

\begin{equation}
p(a,L)=a^{-\tau} f\biggl(\frac{a}{a_{C}}\biggr) \, , 
\label{pda}
\end{equation}

\begin{equation}
\pi(l,L)=l^{-\psi} g\biggl(\frac{l}{l_{C}}\biggr) \, ,
\label{pdl}
\end{equation}

\noindent
where $f(x)$ and $g(x)$ are scaling functions accounting 
for finite size effects 
and $a_{C}$ and $l_{C}$ are the characteristic area and length respectively.
The functions $f(x)$ and $g(x)$ are postulated to have 
the following properties: 
when $x \rightarrow \infty$ they go to zero sufficiently fast to ensure 
normalization; when $x \rightarrow 0$ they tend to a constant, to yield 
simple power law behaviour of the probability distributions in the large size 
limit. This also implies that $\tau$ and $\psi$ are bigger than one.

\noindent
The characteristic area and length are postulated to scale as

\begin{equation}
a_{C} \sim L^{\varphi} \, ,
\label{ac}
\end{equation}

\begin{equation}
l_{C} \sim L^{d_{l}} \, .
\label{lc}
\end{equation}
\noindent

\noindent
In river basins, anisotropies are always present due to a non-zero average 
slope of the landscape and the presence of gravity. Thus, one has to
distinguish between a typical longitudinal length $L$ and a typical 
perpendicular one $L_{\perp}$ (these two length are measured along the two 
principal axis of inertia), which scale as

\begin{equation}
L_{\perp}=L^{H} \, ,
\label{Hurst}
\end{equation}

\noindent
giving $a_{C}\sim L^{1+H}$, i.e. $\varphi= 1+ H$. $H$ is known as the Hurst 
exponent and of course $0\leq H \leq1$. 
In what follow, for the sake of simplicity, we consider basins of square 
shape; the above relations then, refers to dimension of a generic sub-basin
inside the bigger one, whose dimensions are fixed from outside.

 The $ \varphi $ exponent thus 
corresponds to $ \varphi=1+H$ . The $d_{l}$ exponent, characterizing the 
typical length, can be assumed to be the fractal dimension of a stream (for 
fractal river networks, each rivulet going from any site to the outlet is a 
fractal with the same fractal dimension) , and is such that 
$ 1 \leq d_{l} \leq 1+H $. 
The bounds correspond to a straight line and a space-filling behaviour. 
For self-affine river basins we expect $d_{l} = 1 $ and $ H < 1$, 
whereas, when  $ H=1 $ then 
$d_{l} > 1$ in the self-similar case.

For the same quantities, the integrated probability 
distributions can be analogously written:

\begin{equation}
   P(a,L) = a^{1-\tau} F\biggl( \frac{a}{ L^{1+H}}\biggr) \, ,
\label{ipda}
\end{equation}

\begin{equation}
   \Pi(l,L) = l^{1-\psi} G\biggl( \frac{l}{ L^{d_{l}}}\biggr) \, ,
\label{ipdl}
\end{equation}

\noindent
which follow from (\ref{pda}) and (\ref{pdl}) with

\begin{equation}
F(x)=x^{\tau -1} \int^{+\infty}_{x} dy \, \, y^{-\tau} f(y)\,,
\label{F}
\end{equation}

\begin{equation}
G(x)=x^{\psi -1} \int^{+\infty}_{x} dy \, \, y^{-\psi} g(y)\,,
\label{G}
\end{equation}
where sums over variable $y$ have been replaced by integrals in
large $L$ limit.

\noindent
From the above definitions and the properties of $f$ it easily follows  that 

\begin{equation}
\langle a^{n} \rangle \sim L^{(1+H)(n-\tau+1)} \, ,
\label{<an>}
\end{equation}

\noindent
for any $n>\tau \! - \! 1$, while $\langle a^{n} \rangle \sim const.$ if 
$n<\tau \! - \! 1$. Note that both $a$ and $l$ have a lower cutoff 
which is one.

\noindent
This equation, evaluated for $n=1$ gives for the average area,

\begin{equation}
\langle a \rangle \sim L^{(1+H) (2-\tau)} \, .
\label{amedia}
\end{equation}

\noindent
The mean accumulated area $\langle a \rangle$ can be easily shown to be equal
to the distance from a site to the outlet, averaged over all sites. In 
effect, in the sum over all the downstreams (the rivulets going from each 
site to the outlet), the number of times each bond (of unit length) appears 
exactly equals the accumulated area of the associated site. Thus summing 
over all $A_{i}$ is equivalent to a sum over all the downstream lengths: 

\begin{equation}
\langle a \rangle = \langle l_{downstream} \rangle \, . 
\label{addedeq1}
\end{equation}

\noindent
$\langle l_{downstream} \rangle$ can be evaluated replacing in the sum
the distance of each point from the outlet measured along the stream with 
the corresponding euclidean distance $d(x)$ to the power $d_{l}$:

\begin{equation}
\langle l_{downstream} \rangle=\frac{1}{L^{2}}\sum_{x} l_{downstream}(x)=
\frac{1}{L^{2}}\sum_{x} d(x)^{d_{l}} \sim L^{d_{l}}
\label{addedeq2}
\end{equation}

\noindent
This fact is general and the argument we used does not need the knowledge
of the downstream length distribution.
This distribution however can be explicitly derived at least in the
case of {\it directed } networks. We call {\it directed } those networks 
such that each oriented link has positive projection along the diagonal 
oriented towards the outlet. 

\noindent
The reason to introduce this class of networks is that river basins often have 
a {\it quasi-directed} character, due to the fact that they are typically 
grown on a slope which gives a preferred direction to the flow. Moreover
directed trees are much more simple to handle analytically than 
``undirected'' ones.

\noindent
For such trees consider the set of $2L$
diagonals orthogonal to the one passing through the outlet: the downstream 
length is the same for all the points on the same diagonal. Thus the 
number of points at a given distance to the outlet can be easily seen to be:

\begin{equation}
\mbox{number of point at distance} \,\, l =
\left\{
\begin{array}{lll}
l+1     &\,\, & l=1,..L   \\
2L+1-l  &\,\, & l=L+1,..2L
\end{array}
\right.
\label{ups_directed1}
\end{equation}

\noindent
thus the probability density for the downstream lengths has the form
of a power law with exponent $-1$ times a scaling function of the 
argument $l/L$:

\begin{equation}
\pi_{downstream}(l,L)= l^{-1}f_{downstream}\biggl(\frac{l}{L}\biggr)
\label{ups_directed2}
\end{equation}

\noindent
with
\begin{equation}
f_{downstream}(x)=
\min(x^{2},2x-x^{2}) \,\,\,\,\,\,\, 0 \leq x \leq 2
\label{ups_directed3}
\end{equation}

\noindent
The first moment of this distribution again gives eq. (\ref{addedeq2})
with $d_{l}=1$ which is the expected result  for the fractal
dimension of a directed tree.
This result, together with \ref{addedeq2} suggest that downstream length
distribution might have the scaling form:

\begin{equation}
\pi_{downstream}(l,L)=l^{-1}f_{downstream}
\biggl( \frac{l}{L^{d_{l}}} \biggr)
\label{perepeppe}
\end{equation}

\noindent
for the general case.

\noindent
Equations (\ref{addedeq1}) and (\ref{addedeq2}) lead to the following 
expression for the average area:

\begin{equation}
\langle a \rangle \sim L^{d_{l}} \, .
\label{amedia2}
\end{equation}

\noindent
From eq.(\ref{amedia2}) and eq.(\ref{amedia}) we get the scaling relation

\begin{equation}
1+H = \frac{d_{l}}{(2-\tau)} \, .
\label{phi}
\end{equation}

%River basins often have a {\it quasi-directed} character, due to the fact 
%that they are typically grown on a slope which gives a preferred direction to 
%the flow. For {\it directed } networks we will show in Section $4$ that the 
%mean distance to the outlet is equal to $L$, leading to $d_{l}=1$ in 
%equation  (\ref{phi}).

A well known hydrological law, Hack's law \cite{b}, relates the 
length of the longest stream $l$ in the drained area
to the drained area $a$ 
of the basin:

\begin{equation}
l \sim a^{h} \, .
\label{Hack}
\end{equation}

\noindent
The accepted value of $h$ is $h=0.57 \pm 0.06 $ \cite{au,av,az}, whose  
difference from the Euclidean value $0.5$ lead to the first suggestion of the 
fractal nature of rivers \cite{a}.

From equations (\ref{ac}) and (\ref{lc}) it follows that 

\begin{equation}
h=\frac{d_{l}}{1+H} \, .
\end{equation}

Together with $\pi$ and $p$ one can define the conditional probability 
$\tilde{\pi} (l \mid a)$ of finding a main stream with length $l$ in a basin 
with accumulated area $a$. The simplest scenario is that (\ref{Hack}) still
holds and  \cite{i} $\tilde{\pi} (l \mid a)$ is a sharply peaked function of 
$l$ with respect to $a$, i.e. there exists a well-defined constraint between 
lengths and areas,

\begin{equation}
\tilde{\pi} (l \mid a) = \delta(l-a^{h}) \, ,
\label{conditprob}
\end{equation}

\noindent
or more generally \cite{CC},

\begin{equation}
\tilde{\pi}(l\mid a)=l^{-1}\tilde{g}\biggl(\frac{l}{a^{h}}\biggr)\,.
\label{conditprob2}
\end{equation}

\noindent
For the probability density $\pi$, $p$ and $\tilde{\pi}$ the following 
consistency equation must hold:

\begin{equation}
\pi(l,L)=\int^{L^{(1+H)}}_{1} da \,\,\tilde{\pi}(l \mid a) p(a,L)
\label{consistency}
\end{equation}

\noindent
that gives, in the large $L$ limit

\begin{equation}
(\psi-1) d_{l} = (\tau -1 ) (1+H)
\label{sr0}
\end{equation}

\noindent
relating the exponents in the distribution of lengths and in the distribution 
of accumulated areas.

The scaling relations (\ref{phi}) and (\ref{sr0}) can be expressed in a 
simpler form, observing that both $\tau$ and $\psi$ depend on $d_{l}$ and $H$ 
only in the combination $ \frac{d_{l}}{(1+H)} = h $, where $h$ is the 
parameter appearing in Hack's law (\ref{Hack}). Thus

\begin{eqnarray}
\tau & = & 2 - h \, ,
\label{scaling1}   \\
\psi & = & \frac{1}{h} \, .
\label{scaling2}
\end{eqnarray}

\noindent
The exponents characterizing the distributions of accumulated areas and 
upstream length are thus related by the simple expression:

\begin{equation}
\tau = 2 - \frac{1}{\psi} \, .
\label{scaling3}
\end{equation}

For self-affine river basins 

\begin{equation}
H<1 \,\,\, , d_{l}=1 \, ,
\label{selfaffine}
\end{equation}

\noindent
and all exponents can be expressed in terms of the Hurst exponent $H$, giving:

\begin{eqnarray}
\tau & = & \frac{1+2 H}{1+H} \, ,
\label{scaling1sa}  \\
\psi & = & 1+H \, ;
\label{scaling2sa}
\end{eqnarray}

\noindent
while in the self-similar case

\begin{equation}
H=1 \,\,\, , d_{l}>1 \, ,
\label{selfsimilar}
\end{equation}

\noindent
yielding:

\begin{eqnarray}
\tau & = & 2 - \frac{d_{l}}{2} \, ,
\label{scaling1ss} \\
\psi & = & \frac{2}{d_{l}} \, .
\label{scaling2ss}
\end{eqnarray}

\noindent
Note that in both cases, $\tau \leq \! 3/2$. The equality holds only when 
$ H=d_{l}=1$ which corresponds to the mean field situation \cite{aq}.
Likewise,  $h \geq \! 1/2$.

A recently formulated lattice model \cite{ac,ad,ae} based on a minimization 
principle, seems to reproduce quite well the main characteristics of  river 
networks. In this model, a rule for selecting particular configurations in 
the space of spanning trees is given. The ``right" configurations, called 
{\em Optimal Channel Networks} (OCN) are obtained on minimizing a dissipated 
energy, written as:

\begin{equation}
E=\sum_{i} k_i \Delta z(i) Q_i
\label{energy}
\end{equation}

\noindent
where $Q_i$ is the flow rate (the mean annual discharge) in the bond coming 
out from the site $i$, $\Delta z(i)$ is the height drop along the drainage 
direction, and $k_i$ characterizes the local soil properties such as the 
erodability. It will be taken to be equal to one for each bond for 
homogeneous river networks.

\noindent
Given a field of elevations, drainage directions are usually identified by 
steepest descent, i.e. by the largest downward gradient $\nabla z(i)$ of the 
scalar field $z(i)$. This allows us to obtain another expression for the 
dissipated energy on adding the following considerations:

\noindent
\, $i)$ \, in the case of uniform rainfall in time and 
space 

\[
Q_i \sim A_i ,
\]

\noindent
\, $ii)$ experimental observations on  rivers suggest
an empirical relationship between the flow rate and the drop 
in elevation:
\[
\Delta z(i) \sim Q_{i}^{\gamma -1}
\]

\noindent
with a numerical value around $0.5$ for $\gamma$.

\noindent
Thus one obtains, apart from a multiplicative constant, the alternative 
expression of equation (\ref{energy}):

\begin{equation}
E=\sum_{i} k_{i} A_{i}^{\gamma}
\label{E}
\end {equation}

\noindent
which was first proposed by Rinaldo et al. \cite{ac,ad,ae}, and will be 
analyzed further in this paper.

\section{\normalsize\bf The Peano Basin}

The Peano basin  is a deterministic space filling fractal on which exact 
calculations can be carried out \cite{bf}.  It has a spanning tree like 
structure not too dissimilar to that of real basins. The scaling laws for 
such a basin can be obtained exactly and the dissipated energy (\ref{E}) can be
estimated. The latter provides  an upper bound for the minimum energy 
dissipated by an OCN and will be a crucial ingredient for the calculations
to follow.

\noindent
The Peano basin is obtained as follows: start with an oriented link; at step 
2 such a link generates four new links, two resulting from the subdivision in 
half of the old link and preserving its orientation, and the other two having 
a common extreme in the middle point of the old link and both oriented toward 
it (see Fig. 2a). At each successive step, for each link four new oriented 
links are substituted in the way previously described. After $T$ steps the 
fractal has $N_{T} = 4^{T}$ points (excluding the outlet) and it can be mapped 
on a square lattice of size $L=2^{T}$ with bonds connecting first and second 
neighbours to form a spanning tree.

We can associate with each site $i$ of the Peano basin (iterated until step 
$T$) an area $A_{i}(T)$ as in the previously defined lattice model of river 
networks. Let ${\cal V}_{T}$ denote the set of distinct values assumed by the 
variable $A_{i}$ at step $T$. It can be easily checked that ${\cal V}_{T}$
contains ${\cal V}_{T-1}$ and $2^{T-1}$ new distinct values, appearing for 
the first time. Thus  ${\cal V}_{T}$ contains $2^{T}$ distinct numbers. Let 
us call ${\cal A} \doteq \bigcup_{T=0}^{\infty} {\cal V}_{T}$ and $a_{n}$ the 
increasing sequence of numbers in ${\cal A}$ (the distinct values of $A_{i}$ 
that can be generated iterating the construction). For such a sequence, the 
following rule holds

\begin{equation}
a_{n}=3 \, (\sum_{k} c_{k}(n) 4^{k}) +1
\, \, \, \, \, \, \, \, \, n=0,1, \dots
\label{an}
\end{equation}

\noindent
where the $c_{k}(n)$ are the coefficients of the binary expansion of $n$ 
($n=\sum_{k} c_{k}(n) 2^{k}$). In the construction described in Fig. 2b, 
denote by $M_{n}^{T}$ the number of sites $i$ with $A_{i}=a_{n}$ at step $T$. 
The following recursion relation then holds:

\begin{equation}
\left \{
\begin{array}{ll}
M_{n}^{T}=4 \cdot M_{n}^{T-1} -1  & T>t(a_{n})  \\
M_{n}^{T}=1                & T=t(a_{n})  \\
M_{n}^{T}=0                       & T<t(a_{n})
\end{array}
\right.
\label{rec}
\end{equation}

\noindent
where $t(a_{n})$ is the first step in which an area with value $a_{n}$ 
appears, and is given by

\begin{equation}
t(a_{n})=
\left \{
\begin{array}{ll}
0                                             &   n=0             \\
 1+[ \log_2 (n) ]                               &   n>0
\end{array}
\right.
\label{borning}
\end{equation}

\noindent
where [$\cdot$] is the integer part.

\noindent
Solving (\ref{rec}) one gets

\begin{equation}
M_{n}^{T}=
\left \{
\begin{array}{ll}
0 \,\, ,                                      & \,\, T<t(a_{n})             \\
\frac{2}{3} 4^{T-t(a_{n})} +\frac{1}{3} \,\,, & \,\, T \geq t(a_{n})
\end{array}
\right.
\end{equation}

\noindent
and thus all the $a_{n}$ ``born" at the same time step have the same 
probability $p_{\mbox{{\tiny$T$}}}(a_{n}) \doteq 
p(a_{n},L \! = \! 2^{T})=M_{n}^{T}/N_{T}$

Then the integrated distribution of areas $P(A_{i} \! > \! a_{n},
L \! = \! 2^{T})$ assumes a very simple expression for  $a_{n}$ of the form
$4^{t}$ (one can easily check from (\ref{an}) that $a_{2^{t}-1}=4^{t}$),
and is given by

\begin{equation}
P(A_{i} \! > \! a \! = \! 4^{t} \, , \, L \! = \! 2^{T})=
a^{1-\tau} F \biggl( \frac{a}{L^{1+H}}\biggr)
\label{Ppeano}
\end{equation}

\noindent
having  the form (\ref{ipda}) with 

\begin{equation}
\tau=3/2 \, \, , \, \, H=1 
\end{equation}

\noindent
and

\begin{equation}
F(x)=\frac{1}{3} (1-x)\, \, \, \, \, \, \, 0<x<1 .
\end{equation}

\noindent
and $F(x) =0$ when $x> \! 1$.

Equation (\ref{Ppeano}) is obtained on observing that 
$P(A_{i} \! > \! a \! = \! 4^{t} \, , \, L \! = \! 2^{T})=
\sum_{n={2^{t}}}^{2^{T}} p_{\mbox{{\tiny$T$}}}(a_{n})$
depends on $n$ only through $t(a_{n})$, allowing one to replace the sum over 
$n$ with a sum over the steps $s$. Moreover, for each step $s > 0$ there are 
$2^{s-1}$ areas with the same $t(a_{n})=s$.  Thus

\begin{equation}
\begin{array}{l}
P(A_{i} \! > \! a \! = \! 4^{t} \, , \, L \! = \! 2^{T})=
\sum_{s=t+1}^{T} \biggl( \frac{2}{3} 4^{(T-s)}+
\frac{1}{3} \biggr)  \cdot \frac{2^{s-1}}{4^{T}}=  \\
=\frac{1}{3} 2^{-t}(1-2^{2(t-T)})=
\frac{1}{3} a^{-\frac{1}{2}}\biggl( 1- \frac{a}{L^{2}} \biggr)\,.
\end{array}
\end{equation}

\noindent
Similarly, choosing $l$ of the form $l= 2^{t}$ and observing that at step $T$ 
the sites with upstream length greater than or equal to $2^{t}$ are the ones 
in which the accumulated area exceeds $4^{t}$, we find that

\begin{equation}
\Pi(l \geq 2^{t},L=2^{T})= 
l^{1-\psi} G \biggl( \frac{l}{L^{d_{l}}} \biggr)
\end{equation}

\noindent
which is of the form (\ref{ipdl}) with

\begin{equation}
\psi = 2 \, \, , \, \, d_{l} = 1
\end{equation}

\noindent
and

\begin{equation}
G(x)=\frac{1}{3}(1-x^{2}) \, .
\end{equation}

In the following Section we will need some estimates of the mean value of 
$A_{i}^{\gamma}$ for a Peano basin of size $ L= 2^{t}$.

\begin{equation}
\langle A^{ \gamma} 
\rangle =\frac{1}{L^{2}} \sum_{i} A_{i}^{ \gamma}
= \sum_{n=0}^{\infty} 
p_{\mbox{{\tiny$T$}}}(a_{n}) 
{a_{n}}^{\gamma} \, .
\end{equation}

From the expression for $a_{n}$, it  follows that

\begin{equation}
\frac{3}{4} \cdot 4^{t(a_{n})} < a_{n} \leq 4^{t(a_{n})}
\end{equation}

\noindent
($t(a_{n})$ is the ``birth-time" for $a_{n}$), giving

\begin{equation}
\biggl( \frac{3}{4} \biggr)^{\gamma} \alpha (L\! = \! 2^{T})
\leq \, \langle A^{ \gamma} \rangle  \, \leq
\alpha (L\! = \! 2^{T}) \, ,
\label{peanobounds}
\end{equation}

\noindent
where 

\begin{equation}
\alpha (L \! = \! 2^{T}) \doteq
\frac{2}{3}+ \frac {1}{3 4^{T}} + 
\sum_{t=1}^{T} \biggl( \frac {2}{3} 4^{(T-t)}+\frac {1}{3} \biggr)
\frac {2^{t-1}}{4^{T}} 4^{\gamma t} \, .
\label{alpha}
\end{equation} 

\noindent
Performing the summation one gets in the large size limit

\begin{equation}
\alpha \sim
\left \{
\begin{array}{lll}
\frac{1}{3} ( 1+ \frac{1}{1-2^{2 \gamma -1}}) \, & , 
& \,\,\,\, \gamma < 1/2    \\
\frac{1}{3 \log 2} \log L \, & , 
& \,\,\,\,  \gamma = 1/2    \\
\frac{1}{1-2^{1-2\gamma}} L^{2 \gamma-1} \, & , 
& \,\,\,\,  \gamma > 1/2
\end{array}
\right.
\label{alphavacome}
\end{equation}

\noindent
From equation (\ref{peanobounds}), one gets 

\begin{equation}
\langle {A(L)}^{\gamma} \rangle \sim
\left \{
\begin{array}{ll}
const           &         \gamma < 1/2    \\
\log L          &         \gamma = 1/2    \\
L^{2 \gamma-1}  &         \gamma > 1/2
\end{array}
\right.
\label{<a>}
\end{equation}

\noindent
which will be essential to obtain an energy bound for the lattice OCN model.

The scaling exponents for the Peano basin can also be obtained by a 
renormalization group argument. Let us consider, for example, the scaling 
of the accumulated areas. 

\noindent
The self-similar structure of the Peano basin suggests a natural 
``decimation'' procedure. The idea is the following: consider the equations 
relating areas at time step $T$; then, eliminate the variables related to the 
sites that are not present at time step $T-1$. This leads to an effective 
equation describing the same physics on a tree scaled down by a factor $2$.

\noindent
For the sake of simplicity let us consider the Peano basin at the second step 
of iteration. In Fig. 3 let $A_{n}^{(2)}$ denote the variables related to 
sites that are present at step $T=1$ and $B_{n}^{(2)}$ denote the ones that 
will be eliminated by decimation. The upper label refers to the step. In what 
follows it will be useful to write the equations in terms of 
$\tilde{A}_{n}^{(T)}=A_{n}^{(T)}-1$ and $\tilde{B}_{n}^{(T)}=B_{n}^{(T)}-1$.
The areas at step $T=2$ are related by:

\begin{equation}
\left.
\begin{array}{l}
\tilde{A}_{1}^{(2)}=3 \cdot \tilde{B}_{1}^{(2)}+3 \, , \\
\tilde{B}_{1}^{(2)}=\tilde{A}_{0}^{(2)}+2 \cdot \tilde{B}_{0}^{(2)}+3 \, , \\
\tilde{B}_{0}^{(2)}=0 \, .
\end{array}
\right.
\label{rg1}
\end{equation}

\noindent
Elimination of  the $\tilde{B}_{n}^{(2)}$ leads to 

\begin{equation}
\tilde{A}_{1}^{(2)}=3 \cdot \tilde{A}_{0}^{(2)}+ 12 \, .
\label{rg2}
\end{equation}

\noindent
At time step $T=1$ the relation between areas is straightforward: 

\begin{equation}
\tilde{A}_{1}^{(1)}=3 \cdot \tilde{A}_{0}^{(1)} + 3 \, .
\label{rg3}
\end{equation}

\noindent
Equations (\ref{rg2}) and (\ref{rg3}) are the same if

\begin{equation}
\tilde{A}_{n}^{(T+1)}= 4 \, \tilde{A}_{n}^{(T)} \, ,
\label{rg4}
\end{equation}

\noindent
i.e. 

\begin{equation}
(A_{n}^{(T+1)}-1) =  4 \, (A_{n}^{(T)}-1) \, .
\label{rg5}
\end{equation}

\noindent
Denoting by $n^{(T+1)}(a)$ the number of sites with area greater then $a$ at 
step $T+1$, one can easily observe that the number of decimated sites with 
$A>a$ is half of the total number of sites with $A>a$

\begin{equation}
n^{(T+1)}(a)=2 \cdot n^{(T)}(a/4) \, .
\label{rg6}
\end{equation}

Noting that the total number of sites at step $T$, $N_{T}=4^T$, the 
integrated probability $ P(A_{n}^{(T+1)}>a)= \frac{n^{(T)}(a)}{N_{T}}$ 
is given by

\begin{equation}
P(A_{n}^{(T+1)}>a)= b \, P(A_{n}^{(T)}>a/4)
\label{rg7}
\end{equation}

\noindent
with 

\begin{equation}
b=\frac{n^{(T+1)}(a)/4^{T+1}}{n^{(T)}(a)/4^{T}}=\frac{2/4^{T+1}}{1/4^{T}}
=\frac{1}{2} \, .
\label{rg8}
\end{equation}

\noindent
The general solution of equation (\ref{rg7}) is of the form 

\begin{equation}
P(A_{n}^{T}>a) \propto a^{1-\tau}
\label{rg9}
\end{equation}

\noindent
apart from  a superimposed oscillating term given by a periodic function of 
$\log(a)$ with period $1/4$ \cite{bf}. 

\noindent
From equations (\ref{rg7},\ref{rg8},\ref{rg9})

\begin{equation}
\tau = \frac{3}{2} \, .
\label{rg11}
\end{equation}

\noindent
The same argument can be repeated for the distribution of stream lengths, 
recovering the $\psi$ exponent.

\section{\normalsize\bf Analytical Results: homogeneous case}

We now proceed to an analysis of the characteristics of the global minimum of 
the functional $E$ for $\gamma$ in the range $[0,1]$\, for the homogeneous 
case.

Let us consider first  the two limiting cases $\gamma=0$ and  $\gamma=1$. 
If we call $l_{i}$ the weighted length of the stream connecting 
the $i$-th site to the outlet (calculated assigning to each bond a 
length $k_{i}$), it is straightforward to show that

\begin{equation}
\sum_{i} k_{i} A_{i}=\sum_{i} l_{i} \, .
\label{gamma1}
\end {equation}

\noindent
In effect, denoting with $DS(i)$ ($US(i)$) the set of points downstream 
(upstream) with respect to the point $i$ and observing that $A(i)$
equals the number of points in the set $US(i)$ one gets: 
$\sum_{i} l_{i} =\sum_{i}\sum_{j \in DS(i)} k_{j} =
\sum_{j} \sum_{j \in US(i)} k_{j} = \sum_{i} k_{i} A_{i}$ .

\noindent
The minimization of the energy dissipation for $\gamma=1$ thus corresponds to 
the minimization of the weighted path connecting every site to the outlet 
independent of each other.

\noindent
The $\gamma=0$ case, on the other hand, corresponds to the minimization of 
the total weighted length of the tree:

\begin{equation}
E=\sum_{i} k_{i} \, .
\end{equation}

\noindent
In the homogeneous case $k_{i}=1$, $\forall i$, which leads to a high 
degeneracy for both $\gamma = 0$ and $1$.

\noindent
Indeed for $\gamma=0$, each configuration has the same energy (each spanning 
tree on a $L \times L$ square lattice has $L^{2}-1$ links). For $\gamma=1$, 
the minimum of the energy is realized on a large subclass of spanning trees, 
namely all the {\em directed} trees, in which each link has a positive 
projection along the diagonal oriented towards the outlet.

\noindent
For the values of $\gamma \in (0,1)$ there is a competition between both 
mechanisms breaking  the degeneracy and making the search for the global 
minimum  a less trivial problem.

The $\gamma=1$ case gives a minimum energy $E \sim L^{3}$. This can be 
derived observing that all points on a diagonal orthogonal to the one 
passing through the outlet have the same distance from the outlet. Then: 
$ E= \sum_{k=1}^{L-1} k(k+1) + \sum_{k=L}^{2L-2} k(2L-1-k)= L^{2}(L-1)
\sim L^{3}$.  
Thus the value of the energy functional is the same for each {\em 
directed} network and corresponds to the Scheiddeger model of river networks 
\cite{ai} -- all $directed$ trees are equally probable. Such a model can be 
mapped into a model of mass aggregation with injection that has been exactly 
solved by Takayasu et al. \cite{al,am}. The corresponding exponents are:

\begin{equation}
\tau=4/3\,, \psi=3/2\,, H=1/2 \,, d_{l}=1 \,, h=2/3 \,\, .
\label{Sch}
\end{equation}

\noindent
These exponents follow easily from the result $H=1/2$ and from our scaling 
solution of sec. 2. The result $H=1/2$ can be deduced with a simple argument:
since all $directed$ trees are  equally probable, each 
stream behaves like a single random walk in the direction perpendicular to 
the diagonal through the outlet. 
This implies that its perpendicular wandering is $ L_{ \bot} \sim L^{1/2}$. 
Comparing with equation (\ref{Hurst}) one gets $H= 1/2$ .

\noindent
The $\gamma=0$ case leads to  the same energy $E \sim L^{2}$ for each 
network, thus reducing to the problem of random two dimensional spanning 
trees, whose geometrical properties have been calculated on a square lattice 
\cite{ar,as}. The results, in our notation, are:

\begin{equation}
\tau=11/8\,, \psi=8/5\,, H=1 \,, d_{l}=5/4 \,, h=5/8 \,\, .
\label{Dh}
\end{equation}

\noindent
Both (\ref{Sch}) and (\ref{Dh}) are consistent with the scaling relations 
(\ref{scaling1}), (\ref{scaling2}) and (\ref{scaling3}).

We now extend our analysis to the whole range $\gamma \in [0,1]$. We will 
rigorously show \cite{bd} 
that, in the thermodynamic limit, the global minimum in the 
space ${\cal S}$ of all the spanning trees of the functional 
$E(\gamma,{\cal T})=\sum_{i} A_{i}({\cal T})^{\gamma}$ scales as

\begin{equation}
\min_{{\cal T}\in {\cal S}}E(\gamma,{\cal T}) \sim \max(L^{2},L^{1+2\gamma})
\label{Esim}
\end{equation}

\noindent
for all $\gamma \in [0,1]$ .

\noindent
Since $E(\gamma,{\cal T})$ is an increasing function of $\gamma$ and it is 
equal to $L^{2}$ for $\gamma=0$, for $\gamma \geq 0$ it is obvious that

\begin{equation}
E(\gamma,{\cal T}) \geq L^{2} \, .
\label{E>L2}
\end{equation}

\noindent
We now observe that the sum over all the sites can be performed in two steps:

\begin{equation}
E(\gamma,{\cal T})=\sum_{n=1}^{2L-1} 
\sum_{i \in {\cal D}_n} A_{i}({\cal T})^{\gamma}
\label{Ediag}
\end{equation}

\noindent
where ${\cal D}_{n}$ are the lines orthogonal to the diagonal passing through 
the outlet, that we will enumerate  starting from the corner farthest from 
the outlet (see Fig. 4). For  {\em directed} spanning trees one can observe 
that the sum of the areas in a given line ${\cal D}_{n}$ is independent of 
the particular tree, and  is:

\begin{equation}
S_{d}(k)\doteq
\sum_{i \in {\cal D}_{k}} A_{i} =
   \left \{ \begin{array}{ll}
            k (k+1)/2  &  k \leq L  \\
            L^{2}-S_{d}(2L-1-k) & L+1 \leq k \leq 2L-1
            \end{array}
   \right.
\end{equation}

\noindent
where $S_{d}(0) \doteq 0$. This  quantity  only increases on generalizing to 
generic ``undirected" trees and thus for an arbitrary spanning tree,

\begin{equation}
S(k,{\cal T}) \doteq 
\sum_{i \in {\cal D}_{k}} A_{i}({\cal T}) \geq S_{d}(k) \,.
\end{equation}

\noindent
Let us observe that for $k=0,...,(L-1)$ :

\begin{equation}
S(k,{\cal T})+S(2L-1-k,{\cal T}) \geq S_{d}(k)+S_{d}(2L -1-k) = L^2 \, ,
\label{L2}
\end{equation}

\noindent
making it convenient to perform the summation in (\ref{Ediag}) over ``pairs" 
of lines. To get a lower bound for $E$ we need a further inequality:
for every set $\Gamma$ 

\begin{equation}
  \sum_{i \in \Gamma} A_{i}^{\gamma} 
  \geq (\sum_{i \in \Gamma} A_{i})^{\gamma} \, ,
\label{Sw}
\end{equation}
that follows easily from Schwartz inequality,
since $A_{i}\geq 1$ and $0 \leq \gamma \leq 1$.
Now, using (\ref{Ediag}), (\ref{L2}) and (\ref{Sw}) we can write

\begin{eqnarray}
    E(\gamma,{\cal T}) & = & 
\sum_{n=1}^{2L-1} \sum_{i \in{\cal D}_{n}} A_{i}({\cal T})^{\gamma}=
\sum_{n=0}^{L-1} \sum_{    \mbox{{\tiny $   i \in \!\! ({\cal D}_{n}
    \! \cup \! \tilde{{\cal D}}_{n} \! )   $}}  } A_{i}^{\gamma} \geq
\sum_{n=0}^{L-1}(\sum_{    \mbox{{\tiny $   i \in \!\! ({\cal D}_{n} 
    \! \cup \! \tilde{{\cal D}}_{n} \! )   $}}  } A_{i}({\cal T}))^{\gamma}  
    \nonumber \\
     & = & \sum_{n=0}^{L-1}[S(n,{\cal T})+S(2L-1-n,{\cal T})]^{\gamma}  \geq
	\sum_{n=0}^{L-1} L^{2\gamma} = L^{1+2 \gamma } \, ,
\end{eqnarray}

\noindent
where $\tilde{{\cal D}}_{n}= {\cal D}_{(2L-1-n)}$. The equality in the last inequality holds for {\em directed} networks. We can thus write

\begin{equation}
    E(\gamma,{\cal T}) \geq L^{1+2\gamma} \, .
\label{E>L1+2g}
\end{equation}

\noindent
Equation (\ref{E>L1+2g}) together with the (\ref{E>L2}) gives the lower bound

\begin{equation}
E(\gamma,{\cal T}) \geq \max(L^{2},L^{1+2\gamma}) \, ,
\label{ineq1}
\end{equation}

\noindent
that holds for every ${\cal T} \in {\cal S}$ and thus also for the minimum 
over ${\cal T}$. Using the results of the previous Section, we can exhibit a 
tree on which the bound is realized. In effect, the Peano network can be 
mapped on a square lattice only considering links between first and second 
nearest neighbours, but (\ref{ineq1}) can be analogously obtained for such a 
lattice on rearranging in an opportune way the summation in equation (\ref
{Ediag}). If we call ${\cal T}_{P}$ the spanning tree given by the Peano basin
we know from equation (\ref{<a>}) that, except for logarithmic corrections 
for $\gamma=1/2$, $E(\gamma,{\cal T}_{P}) \sim \max(L^{2},L^{1+2\gamma})$. Thus

\begin{equation}
\min_{{\cal T} \in {\cal S}}E(\gamma,{\cal T}) \leq E(\gamma,{\cal T}_{P}) \, ,
\end{equation}

\noindent
and we get equation (\ref{Esim}).

We now proceed to the calculation of the scaling exponents.

\noindent
For a directed path, from equation (\ref{gamma1}),

\begin{equation}
\langle a \rangle \sim L \, .
\end{equation}

\noindent
For generical ``undirected" networks, let us write as in (\ref{amedia2}) 

\begin{equation}
\langle a \rangle \sim L^{d_{l}}
\end{equation}

\noindent
($d_{l}$ could be be somewhat bigger than one if one assumes a 
``quasi-directed" behaviour).

\noindent
Using  eq.  (\ref{<an>}) with $n=\gamma$ and the above result on the scaling 
of energy: $ E= L^{2} \langle a^{\gamma} \rangle \sim L^{1+2 \gamma}$, one  
gets

\begin{equation}
2 \gamma -1 =(1+H)( \gamma - \tau +1) 
\label{sr2}
\end{equation}

\noindent
holding for $\gamma > \tau \! - \! 1$ .

\noindent
Equations (\ref{sr2}) together with the scaling relation  (\ref{phi}) can be 
solved with respect to $\tau$ and $H$, and gives, for $ \gamma>1/2$

\begin{equation}
\left \{
\begin{array}{l}
\tau=\frac{3(1-\gamma) + (d_{l}-1) (1+\gamma)}
        {2(1-\gamma)+ d_{l}-1}
\\
H=\frac{d_{l}-\gamma}{1-\gamma}
\end{array}
\right.
\label{tpe}
\end{equation}

\noindent
(the constraint $\gamma> \tau \! - \! 1$ become $\gamma>1/2$, independently 
of $d_{l}$). Thus, if $H \leq 1$, for any $\gamma<1$ $d_{l}=1$, yielding 

\begin{equation}
\tau = 3/2 \,\,,\,\, H=1 \,\,,\,\, \psi=2 \,\,,\,\, d_{l}=1 \,\,,\,\, h=1/2
\label{tp}
\end{equation}

\noindent
for $\gamma \in (1/2,1)$.
The exponents are the same as in the mean field theory \cite{al,am} of 
the Scheidegger model and the same as in the Peano case \cite{footnote}.

\noindent
Note that equations (\ref{tpe}) are meaningless for $\gamma=1$, in which case 
(\ref{tp}) does not hold, consistent with the Scheiddeger results (\ref{Sch}).

\noindent
When $ 0 < \gamma < 1/2$, all we can say is that $ \tau > 1+ \gamma$. However 
if $d_{l} = 1$ for any $ \gamma \in (0,1)$ (i.e. the Optimal Channel is 
$quasi-directed$) then one might expect that $H=1$ for all values of $\gamma$ 
and thus eq. (\ref{tp}) holds in the whole range $\gamma \in (0,1)$.
This prediction is confirmed by results of numerical simulations.

\section{\normalsize\bf Analytical Results: heterogeneous case}

In this Section we focus our attention on the case in which some sort of 
quenched disorder is present in the basin. Two cases have been analyzed:

{\em i}) $\,$random bonds, modelling heterogeneity in the local properties 
of the soil;

{\em ii}) random injection, modelling non-uniformity in the rainfall.

In the first case, we will show that the energy can be bounded from above 
with the corresponding one in the absence of disorder. This will give in the 
large $L$ limit an upper bound for the $\tau$ exponent that we will see is 
realized for all the $\gamma \in (1/2,1)$. In the case of random rainfall, 
we will show that this kind of randomness does not affect the scaling 
behavior of the dissipated energy  in the large $L$ limit. All the analytical 
results found in Section $4$ for the homogeneous case, being based on the 
energy estimate in the thermodynamic limit, can thus be extended to that case, 
giving the same values for the exponents.

In the case of random bonds, we associate with each bond of the $L \times L $ 
square lattice ($2L(L-1)$ bonds) a quenched random variable $k_{b}$, 
arbitrarily distributed  such that $\langle k_{b} \rangle =1$. The label $b$ 
ranges over all the bonds of the lattice. The $2L(L-1)$ variables are chosen 
independent of each other and identically distributed. 

In what follows $b \, (i,{\cal T})$ will denote the label associated with the 
bond coming out from the site $i$ on the tree ${\cal T}$. Let 
${\cal T}^{*}(\gamma)$ denote one of the trees on which the minimum of the 
energy $E(\gamma, {\cal T})$ is realized in the homogeneous case for a given 
value of $\gamma$, and {\cal S} the set of all the spanning trees: 

\begin{equation}
E(\gamma) \doteq \min_{{\cal T} \in {\cal S}} E(\gamma, {\cal T})=
\sum_{i} A_{i}({\cal T})^{\gamma} 
= \sum_{i} {A_{i}({\cal T}^{*}(\gamma))}^{\gamma} \, .
\label{homog}
\end{equation}

Denoting by $ E_{D}(\gamma) $ the minimum of the energy in the heterogeneous 
case, averaged over the quenched disorder, the following inequality holds:

\begin{eqnarray}
    E_{D}(\gamma) & \doteq &
\langle \min_{{\cal T} \in {\cal S}}
\sum_{i} k_{b(i,{\cal T})} A_{i}({\cal T})^{\gamma} \rangle 
\leq \langle \sum_{i} k_{b(i,{\cal T}^{*}(\gamma))} 
{A_{i}({\cal T}^{*}(\gamma))}^{\gamma} \rangle \nonumber \\
                  &  =     &
\langle \sum_{i} k_{b(i,{\cal T}^{*}(\gamma))} \rangle 
A_{i}({\cal T}^{*}(\gamma))^{\gamma} = 
\sum_{i} {A_{i}({\cal T}^{*}(\gamma))}^{\gamma} = E(\gamma) \, .
\label{energiarb}
\end{eqnarray}

\noindent
The energy in the presence of this kind of disorder is thus bounded from 
above by the energy in the absence of disorder for any value of the $\gamma$ 
parameter. In the large size limit this result gives bounds on the scaling 
exponents. Equation (\ref{<an>}) evaluated for  $ n=\gamma$ gives 

\begin{equation}
\langle a^{\gamma} \rangle \sim L^{(1+H_{D})(\gamma - \tau +1 )}
\label{agammam}
\end{equation}

\noindent
which holds for any $\gamma > \tau_{D} -1=1-h_{D} $, thus at least for any 
$\gamma > 1/2$ since $h_{D} \geq 1/2 $. Here and in what follows variables 
with the $D$ index refer to the random bond case.
%%%%%%%%%%%%%%%%%%%%%%%%%%%%%%%%%%%%%%%%%%%%%%%%%%%%%
Equation (\ref{agammam}), compared with equation (\ref{Esim}) leads to

\begin{equation}
(1+H_{D}) (\gamma - \tau_{D} +1) + 2 \leq 
1+2 \gamma \, \, \, \, ,   1/2          \! \! \leq \gamma \leq \! \!  1.  
\label{rbexp}
\end{equation}

In the case of self-affine behaviour, using eq. (\ref{phi}), equation 
(\ref{rbexp}) gives:  

\begin{equation}
H_{D} \geq 
1  \,\,\, \, ,   1/2   \! \! \leq \gamma <   \! \!  1.  
\label{rbexpsa2}
\end{equation}

\noindent
As before, for the $ \gamma = 1 $ case the above inequalities are useless.
In the case of self similar behaviour, inequalities for the fractal dimension 
$d_{l,D}$ can analogously be deduced:

\begin{equation}
d_{l,D} \leq 
1  \,\,   \, \, , 1/2  \! \! \leq \gamma \leq \! \!  1  
\label{rbexpss1}
\end{equation}

\noindent
Because, in general $0 \leq H \leq 1$ and $ 1 \leq d_{l} \leq 1+H $ ,
equations (\ref{rbexpsa2}) and (\ref{rbexpss1})  together yield for 
$\gamma \in (1/2,1)$:

\noindent
$H_{D}=1$, $ d_{l,D}=1 $, i.e. the mean field values. 

\noindent
The $\gamma=1$ case has been exactly solved \cite{aq}, and it can be mapped 
into a problem of a directed polymer in a random medium or equivalently a 
domain wall in a disordered two dimensional ferromagnet\cite{an,ao,ap}, for 
which an exact solution is known.
 
The corresponding values for the exponents are:

\begin{equation}
\tau= 7/5 \, \, , \, \, \psi=5/3 \, \, , \, \, H=2/3 \, \, ,
\, \, d_{l}=1 \, \, , h=3/5 \, .
\label{rbgamma1}
\end{equation}

Disorder can be introduced in the system in another way, replacing the 
constant injection in each site of the lattice with a random, quenched, local 
injection, i.e. a spatial inhomogeneity in the rainfall $r_{i}$ in eq. 
(\ref{aree}). 

\noindent
In order to do that, one can associate with each site $i$ of the lattice a 
random variable $r_{i}$. The variables are chosen to be independent of each 
other, identically distributed and with mean $\langle r_{i} \rangle =1 $ .

\noindent
The accumulated areas must then satisfy, as in (\ref{aree})

\begin{equation}
A_{i}= \sum_{j } w_{i,j} A_{j} + r_{i} \, ,
\label{areerandom}
\end{equation}

\noindent
in such a way that 
\begin{equation}
\, \, A_{i}=\sum_{j} \lambda_{i,j} r_{j} \,\,\, , \,\,\,\,\,\,\,\,
\mbox{with}  \,\,\, \lambda_{i,j} \doteq 
\left \{
\begin{array}{ll}
1  &  \,\,\,\mbox{if $i$ is connected to $j$         } \\
   &  \,\,\,\mbox{through upstream drainage directions} \\
   &  \,\,\,\mbox{or if $j=i$ ,                      } \\
0  &  \,\,\,\mbox{otherwise .                        }
\end{array}
\right.
\label{areerandom2}
\end{equation}

\noindent
The minimum of the energy averaged over the ``random-rainfall'' will be 
denoted by $E_{rr}(\gamma)$, and for a given value of $\gamma$ is given by 

\begin{equation}
E_{rr}(\gamma) \doteq \langle \min_{{\cal T} \in {\cal S}} 
\sum_{i} A_{i}(\{r_{j}\},{\cal T})^{\gamma} \rangle \, ,
\label{energiarr}
\end{equation}

\noindent
where ${\cal S}$ denotes the set of all spanning trees ${\cal T}$ and 
$\{r_{j}\}$ denotes the whole set of random variables. As in (\ref{homog}) 
${\cal T}^{*}$ will be one of the trees for which the minimum of the energy 
is realized in the absence of randomness in the rainfall and for a given 
value of $\gamma$. Then

\begin{eqnarray}
    E_{rr}(\gamma) & \doteq &
                       \langle \min_{{\cal T} \in {\cal S}} \sum_{i} 
                       A_{i}(\{r_{i}\},{\cal T})^{\gamma} \rangle 
                             \leq 
                         \langle \sum_{i} {A_{i}
                         (\{r_{i}\},{\cal T}^{*}(\gamma))}^{\gamma} \rangle 
			 \nonumber \\
                   &  =     &
                         \langle {\sum_{i} \biggl( \sum_{j} \lambda_{i,j}({
                         \cal T}^{*}(\gamma)) r_{j} \biggr)}^{\gamma} \rangle 
                            \leq 
                         \sum_{i} \sum_{j} \lambda_{i,j}({
                         \cal T}^{*}(\gamma)){\langle r_{j}\rangle}^{\gamma} 
			 \nonumber \\
                   &  =     & \sum_{i} {A_{i}({\cal T}^{*})}^{\gamma}
                             =
                         E(\gamma) \, .
\label{2energiarr}
\end{eqnarray}

\noindent
Thus 

\begin{equation}
E_{rr}(\gamma) \leq E(\gamma) \sim \min (L^{2},L^{1+2 \gamma}) \, .
\label{ubenergiarr}
\end{equation}

\noindent
In this case, it is possible to bound the energy also from below with an 
argument analogous to that used in Section $4$ for the homogeneous case. 
The detailed calculation is given in the appendix. Thus one can conclude 
that

\begin{equation}
E_{rr}(\gamma) \sim \min(L^{2}, L^{1+2 \gamma}) \, ,
\label{enerrrsim}
\end{equation}

\noindent
and all the results of Section $4$ hold.

\section{\normalsize\bf Numerical Results: global minimum}

We have carried out extensive numerical investigations of OCN along two 
avenues:

{\em i}) $\,$the search for the global minimum with a Metropolis algorithm
for $\gamma = 1/2$. 

{\em ii}) the statistics over local minima for $\gamma=1/2$; strikingly these 
yield consistent but  different values for the scaling exponents. By a local 
minimum, we mean a configuration (a spanning tree) of the network such that 
no link can be changed without increasing the energy. The global minimum is 
of course also a local minimum; but in the two cases we found different
statistics, that is suggestive of a very rich structure of the 
energy-landscape. We will postpone the results concerning the latter subject
to the next Section, focusing  our attention only on the scaling properties 
of the global minimum.

In the computer simulations we considered a square lattice with all sites on 
one side allowed to be an outlet fot the network. Periodic boundary 
conditions were chosen in the other direction.

\noindent
Once the simulation has been performed over the whole lattice, the basin with 
the biggest drained area is selected and only sites contained therein are 
used to calculate statistical quantities. Multiple outlets are allowed in 
order to minimize finite-size effects. The rainfall is assumed to be uniform 
over the whole lattice. The optimization method used is  simulated annealing,
in which  a parameter $T$ analogous to the temperature is introduced and 
lowered during the simulation. For each $T$ value the system is relaxed in 
the following way: a new allowed configuration ``near"  the initial one 
(``near" means one differing from the previous one only in one link) is 
randomly chosen; the dissipation energy of the new configuration is 
calculated and compared with the value of the old one. The new configuration 
is accepted with probability $1$ if $\Delta E$ is negative, and with 
probability $\exp[- \frac{\Delta E}{T}]$ otherwise.

In short a sketch of the algorithm is:

\noindent
{\em i}) $\,\,$ Generation of a random initial configuration:
		\begin{quotation}
		\noindent
		Starting from a given, fixed tree
		we obtain a random initial configuration
		running steps {\em ii}) and {\em iii})
		several times. In this way
		only changes preserving the spanning loopless
		structure are accepted.
		\end{quotation}
{\em ii}) $\,$  Random changes of the configuration:
		\begin{quotation}
		\noindent
                We select randomly one site $i$ and then again
                randomly one of the ``free" (not yet belonging
		to the tree) links connected with $i$, if any.
		\end{quotation}
{\em iii})      Geometrical controls:
		\begin{quotation}
		\noindent
	        The absence of a loop in the new configuration
		is checked. If a loop is present, step {\em ii}) 
		is repeated.
		\end{quotation}
{\em iv}) $\,$  Energetic control:
		\begin{quotation}
		\noindent
		The change,  $\Delta E$, in the dissipation energy
		dissipation is calculated. If it is
		negative we go on to step {\em v}). Otherwise,
		a random number $p$ uniformly
		distributed in the interval $[0,1]$ is
		generated and compared with $\exp[-\frac{\Delta E}{T}]$:
		if $\exp[-\frac{\Delta E}{T}] \leq p$
		we go on to step {\em v}) , otherwise the change is
		rejected and we go back to step {\em ii}).
		\end{quotation}
{\em v}) $\,\,$ Recalculation of changed quantities:
		\begin{quotation}
		\noindent
		All variables involved in the change
		are updated.
		\end{quotation}
{\em vi})$\,$   Lowering of the $T$ parameter:
		\begin{quotation}
		\noindent
		In each cycle, the $T$ parameter is decreased
		with the following rule: at the $ n$-th cycle
		$T$ is given by $T(n)=\alpha ^{n}T(0)$, where
		$\alpha$ is a parameter very close to $1$ (we choose
		$\alpha = 0.986$) and $T(0)$ is a suitable 
		chosen constant.
		\end{quotation}
After step {\em i}), steps {\em ii})-{\em v}) are repeated many times, say 
$N$. Then we go to step {\em vi}) in which the $T$ parameter is lowered and 
the entire algorithm (except step {\em i}) ) is repeated. The annealing 
process stops when $T$ reaches very low values ($ \approx 10^{-4}$).

%%%%%%%%%%%%%%%%
%Simulations have been performed varying the values $\gamma= 0.1; 0.2; 0.3; 
%0.4 $ for five different initial configurations. The measured
%statistical quantities $\tau$ and $\psi$ gave the values $\tau =.. \pm .. $ 
%and $ \psi=.. \pm ..$ [errore.....] and seems to be independent from the 
%initial configuration (i.e. in the five runs).
%The $\gamma = 0.5$ case have been studied more intensively.
%%%%%%%%%%%%%%%%%

The simulations have been repeated varying the initial configuration for
a size $L= 128$. The statistical  quantities do not depend on the initial
data. The integrated probability distributions for the accumulated areas and 
mainstream lengths averaged over 10 trials are shown in Fig.5 and Fig.6 and 
give $\tau =1.50 \pm 0.02 $ and $\psi=2.00 \pm 0.02$ where the error is 
estimated as the root mean square root over the ten trials. 

\noindent
These results are in perfect agreement with equation (\ref{scaling3})
and confirm that the analytical results hold when $\gamma = 1/2 $.

%%%%%%%%%%%
%All the above numerical results seems to confirm
%that analytical results can be extended giving
%mean field behaviour 
%in the whole $(0,1)$ region.
%%%%%%%%%%%

\section{\normalsize\bf Numerical Results: local minima}

Homogeneous OCN yield results in good agreement with experimental data on 
rivers when statistics based on local minima is calculated. This suggested 
\cite{at} the concept of feasible optimality according to which nature is 
``unable" to reach the true ground state when complex systems are involved in 
optimization problems. The optimization just stops when one of the local 
minima is reached. Within this framework, the scaling properties of real
rivers should be reproduced considering the ensemble of local minima.

We have performed the search for local minima with an algorithm equivalent to 
a $T=0$ Metropolis scheme, i.e. one in which new configurations are accepted 
only if energetically favourable. The simulation has been repeated $40$ 
times, starting with different, randomly chosen, initial data and varying the 
size $L$ of the system: $ L=32, 64, 128$ .

The values obtained for the characteristic exponents of the probability
distribution considered above, $ \tau$ and $\psi$ are in very good agreement 
with experimental data. The distributions obtained starting from different 
initial condition do not substantially differ from one another, all yielding 
the same value for the exponents.

The statistical quantities thus seem to be robust with respect to variations 
of initial conditions, showing the self-averaging of the scaling parameters.

Results obtained averaging over 40 local minima are shown in Figs.7 and 8
and give $\tau = 1.45 \pm 0.02$ and $\psi = 1.82  \pm 0.02$. The results are 
consistent with the scaling relations. A collapse test has been done to 
verify the consistency of numerical values of the exponents with the finite 
size scaling hypothesis (Fig.9).

\section{\normalsize\bf Summary and Conclusions}

In this paper,  we have studied the Optimal Channel Network for the drainage 
basin of a river.

\noindent
Within the framework of the finite size scaling hypothesis for the 
distributions of accumulated areas and mainstream lengths, we deduced the 
exact scaling behaviour for the tree (drainage network) for which the 
absolute minimum of dissipated energy is realized in both  the homogeneous 
case and in the presence of randomness. The scaling exponents in the 
homogeneous case are found to be the mean field ones and  differ from those 
measured in real rivers.

\noindent
Numerical results were obtained  both for  the statistics of the global 
minimum (confirming the analytical predictions) and for local minima. The 
statistical exponents characterizing the local minima definitely differ from 
the mean field ones. They seems to be in a new distinct universality class 
and in agreement with data from real rivers. This suggests  that real rivers, 
during their evolution, do not visit all of configuration space but are soon 
trapped in a metastable state, i.e. a local minima of the dissipated energy.

%\noindent
%Lot of work still needs to be done. In the context of OCN a complete
%description of local minima is still lacking. A wider approach to the 
%description of river networks should include the mechanisms leading to 
%aggregation and driving rivers evolution.

\noindent
The authors wish to thank A. Rinaldo and M. Cieplak for many enlightening 
discussions.
This work was supported by NASA, NATO, NSF, The Petroleum Research Fund
administered by the American Chemical Society and The Center for Academic
Computing at Penn State.
\section{Appendix}

Using the  notation of Section $4$ we denote by ${\cal D}_{n}$ the lines 
orthogonal to the diagonal passing through the outlet, enumerated starting 
from the corner farthest from the outlet with $0$. It will be useful to 
associate  each line ${\cal D}_{n}$ with $n \leq L-1$ with a 
$\tilde{{\cal D}}_{n}$:

\begin{equation}
\tilde{{\cal D}}_{n}= {\cal D}_{(2L-1-n)} \, .
\label{app1}
\end{equation}

\noindent
In what follows we will choose the $\{r_{i}\}$ to be independent random 
variables with values in the interval $[ 0, 2 ]$ and such that 
$\langle r_{i} \rangle = 1$. We then proceed to the first step of the proof 
as in Section 4, observing that the sum over all the sites in equation 
(\ref{energiarr}) defining the energy can be performed in two steps:

\begin{eqnarray}
E_{rr}(\gamma) & \geq & \langle \min_{{\cal T} \in {\cal S}}
   \sum_{i} A_{i}(\{ r_{i} \} ,{\cal T})^{\gamma} \rangle  \nonumber   \\
 & \geq & \langle \min_{{\cal T} \in {\cal S}}
   \sum_{n=0}^{L-1} \biggl( \sum_{    
   \mbox{{\tiny $   i \in \!\! ({\cal D}_{n}
   \! \cup \! \tilde{{\cal D}}_{n} \! )   $}}    } 
   A_{i}( \{ r_{i} \},{\cal T}) \biggl) ^{\gamma}   \rangle   \nonumber  \\
 &  \geq &  
\sum_{n=0}^{L-1} \langle \biggl(
   \sum_{k=1}^{n}      \sum_{\mbox{{\tiny $i \in \!\! {\cal D}_{k}$}}} r_{i}+ 
   \sum_{k=1}^{2L-1-n} \sum_{\mbox{{\tiny $i \in \!\! {\cal D}_{k}$}}} r_{i}
   \biggr)^{\gamma} \rangle  \, \, ,
\label{app2}
\end{eqnarray}

\noindent
where in the last step the equality holds only for directed trees. 

\noindent
The last expression can be written in a more convenient form introducing 

\begin{equation}
\mu_{i} =
\left \{
\begin{array}{ll}
2 & \,\, \,\, \,\, i \in {\cal D}_{k} \,,\,\mbox{ with } k \leq n  \\
1 & \,\, \,\, \,\, \mbox{otherwise} \,\,.
\end{array}
\right.
\label{app3}
\end{equation}

\noindent
Then

\begin{eqnarray}
E_{rr} &   \geq   & \sum_{n=0}^{L-1} 2^{\gamma} L^{2\gamma} \langle
   \biggl( \frac {
   \sum_{k=1}^{2L-1-n} \sum _{\mbox{{\tiny $i \in \!\! {\cal D}_{k}$}}} 
   \mu_{i} r_{i}
   }{2 L^{2}} \biggr) ^{\gamma} \rangle  \nonumber \\
       & \geq & \sum_{n=0}^{L-1} 2^{\gamma} L^{2\gamma} 
    \frac {
   \sum_{k=1}^{2L-1-n} \sum _{\mbox{{\tiny $i \in \!\! {\cal D}_{k}$}}} 
   \mu_{i} \langle r_{i} \rangle 
   }{2 L^{2}}   = 2^{\gamma} L^{1+2 \gamma} \, \, ,
\label{app4}
\end{eqnarray}

\noindent
where the last inequality follows on observing that 

\[
 \sum_{k=1}^{2L-1-n} \sum _{\mbox{{\tiny $i \in \!\! {\cal D}_{k}$}}} 
   \mu_{i} = L^{2}  .
\]

\noindent
Thus,  for $r_{i} \leq 2 $, the argument between the brackets is less than or 
equal to one. Furthermore, for any $x$ and $  \gamma \in [0,1]$,  
$x^{\gamma} \geq x $. 

\noindent
Equation (\ref{app4}) together with equation (\ref{ubenergiarr}) of Section 5 
gives

\begin{equation}
2^{\gamma} L^{1+2\gamma} \leq E_{rr}(\gamma) \leq E(\gamma) \sim L^{1+2\gamma}
\label{app5}
\end{equation}

\noindent
and thus

\begin{equation}
E_{rr}(\gamma) \sim L^{1+2\gamma}  .
\label{app6}
\end{equation}

\newpage

\begin{figure}
\caption{(1.a) The basin of Fella River in the north-east of Italy
reconstructed from a Digital Elevation Map.
(1.b) A lattice river basin of size $L=5$. In each site $i$ the
value of the cumulated area $A_{i}$  is displayed. The darkest line 
represent the main stream of the entire basin.}
\end{figure}

\begin{figure}
\caption{(2.a) The Peano basin at iteration step $ T=0$, $T=1$,
$ T=2$ and $T=3$, with
the accumulated areas displayed.
(2.b) To obtain Peano basin at time step $T+1$, 
one  keeps the basin at time step $T$, cuts the outlet and 
joins four copies of what is obtained as 
illustrated in the figure. In
this way, the recursion relation 
can be easily understood.}
\end{figure}

\begin{figure}
\caption{The renormalization group argument for the Peano basin. 
B-sites die under decimation.}
\end{figure}

\begin{figure}
\caption{Each dashed line divides the lattice in two parts.
$\sum_{i \in \cal{D}} A_{i} \,$ is at least equal at the number
of sites contained in the part of the lattice with border $\cal{D}$
and without the outlet.}
\end{figure}

\begin{figure}
\caption{The distribution $P(A>a)$ versus a for a basin of linear
size $L=128$. Such a distribution has been obtained by mean of a 
Metropolis-like algorithm (attempting to seek the 
global minimum) and averaged over ten samples.
The slope displayed is $1-\tau=-0.50$.}
\end{figure}

\begin{figure}
\caption{The distribution $\Pi(l>\lambda)$ versus $l_{0}$ for the same 
conditions as in Fig. 5. The slope $1-\psi=-1.00$}
\end{figure}

\begin{figure}
\caption{$P(A>a)$ versus $a$ for three different sizes of the basin:
$L=32, L=64$ and $L=128$.  Local minima are considered here, the distribution
is obteined averaging over $40$ samples.
The slope is $1-\tau =-0.45 $}
\end{figure}

\begin{figure}
\caption{The same as in Fig. 7 for the upstream length distribution.
The slope is $1-\psi= -0.82$}
\end{figure}

\begin{figure}
\caption{The result of the collapse test for the accumulated 
area distribution in the case of the local minima.}
\end{figure}


\newpage

\begin{thebibliography}{99}

\bibitem{a} 
B. B. Mandelbrot, {\it The Fractal Geometry of Nature}
( Freeman, New York, 1983 ).

\bibitem{d}  D. G.
Tarboton, R. L. Bras and I. Rodriguez-Iturbe, {\it Water Resour. Res.}
{\bf 24}, 1317 (1988);{\bf 25}, 2037 (1989); {\bf 26}, 2243 (1990).

\bibitem{n}
P. La Barbera and R.
Rosso, {\it Water Resour. Res.} {\bf 25}, 735 (1989).

\bibitem{p}
D. R. Montgomery and W. E. Dietrich,
{\it Nature} {\bf 336}, 232 (1988).

\bibitem{ai}
A. E. Scheidegger, {\it Bull. Assoc. Sci. Hydrol.}
{\bf 12}, 15 (1967).

\bibitem{f}
P. Meakin, J. Feder and T. J$\o$ssang, {\it Physica A}
{\bf 176}, 409 (1991).

\bibitem{ba}
G. R. Willgoose, R. L. Bras and I. Rodriguez-Iturbe,
{\it Water Reoour. Res.}{\bf 27}, 1671 (1991); {\bf 27}, 1685 (1991);
{\it Geomorphology }{\bf 5}, 21 (1992).

\bibitem{bb}
S. Kramer and M. Marder,
{\it Phys. Rev. Lett.} {\bf 68}, 205 (1992).

\bibitem{bc}
R. L. Leheny and S. R. Nagel,
{\it Phys. Rev. Lett.} {\bf 71}, 1470 (1993).
 
\bibitem{AA}
A. Giacometti, A. Maritan and J. R. Banavar
{\it Phys. Rev. Lett.} {\bf 75}, 577 (1995).

\bibitem{ac}
I. Rodriguez-Iturbe, A. Rinaldo, R. Rigon, R. L. Bras, E. Ijjasz-Vasquez,
{\it Water Resour. Res.} {\bf 28}, 1095 (1992).

\bibitem{ad}
I. Rodriguez-Iturbe, A. Rinaldo, R. Rigon, R. L. Bras, E. Ijjasz-Vasquez,
{\it Geophys. Res. Lett.} {\bf 19} 889 (1992).

\bibitem{BB}
K. Sinclair and R. C. Ball
{\it Phys. Rev. Lett.} {\bf 76}, 3360 (1996)

\bibitem{h}
 T. Sun, P. Meakin and T. J\o ssang, {\it Phys. Rev. E}
{\bf 49}, 4865 (1994).

\bibitem{ag}
T. Sun, P. Meakin and T. J\o ssang, {\it Phys. Rev. E}
{\bf 51}, 5353 (1995).

\bibitem{ah}
T. Sun, P. Meakin and T. J\o ssang,
{\it Water Resour. Res.}  {\bf 30}, 2599 (1994).

\bibitem{i}
A. Maritan, A. Rinaldo, R. Rigon, A. Giacometti and I. Rodriguez-Iturbe,
{\it Phys. Rev. E} {\bf 53}, 1510 (1996)

\bibitem{b} 
J. T. Hack, {\it U.S.  Geol. Surr. Prof.} 
Paper {\bf 294-B}, 1 (1957); {\bf 504-B}, 1 (1965).

\bibitem{au}
D. M. Gray,
{\it J. Geophys. Res.} {\bf 66}, 1215 (1961).

\bibitem{av}
W. B. Langbein,
{\it U.S. Geol. Surv. Prof. Paper} {\bf 968-C}, 1 (1947).

\bibitem{az}
J. E. Muller,
{\it Geol. Soc. A Bull.}{\bf 84},3127 (1973).

\bibitem{CC}
R. Rigon, I. Rodriguez-Iturbe, A. Rinaldo, A. Maritan, A. Giacometti
to appear in {\it Water Resour. Res.}

\bibitem{aq}
A. Maritan, F. Colaiori, A. Flammini, M. Cieplak and J. R. Banavar,
{\it Science} {\bf 272}, 984 (1996).

\bibitem{ae}
A. Rinaldo {\it et. al.},
{\it Water Resour. Res.} {\bf 28}, 2183 (1992).

\bibitem{bf}
A. Flammini and F. Colaiori,
submitted to {\it Journal of Physics A}.

\bibitem{al}
H. Takayasu, M. Takayasu, A. Provata and G. Huber,
{\it J. Stat. Phys.} {\bf 65}, 725 (1991).

\bibitem{am}
M. Takayasu, H. Takayasu, Y. -H Taguchi,
{\it Int. J. Mod. Phys. B} {\bf 8}, 3887 (1994).

\bibitem{ar}
S. S. Manna, D. Dhar and S. N. Majumdar,
{\it Phys. Rev. A } {\bf 46}, R4471 (1992).

\bibitem{as}
A. Coniglio,
{\it Phys. Rew. Lett.} {\bf 62}, 3054 (1989).

\bibitem{bd}
 A. Rinaldo, A. Maritan, F. Colaiori, A. Flammini,  R. Rigon,
 I. Rodriguez-Iturbe and J. R. Banavar,
{\it Phys. Rev. Lett.} {\bf 76}, 3364 (1996).

\bibitem{footnote}
OCN networks on a square lattice have also been studied by Sun, Meakin and
J\o ssang \cite{h}. 
They allowed each site of the border to be an outlet 
and considered the distribution of accumulated areas of a site belonging to 
the border (and not for a generic site of the basin as we have been doing in 
the present paper). They found a power law behaviour for such a distribution:
$
P_{\partial}(a) \sim a^{- \tau_{\partial}}
$
with $\tau_{\partial}$ numerically estimated to be $\simeq 1.51$. If finite 
size scaling is invoked for such a distribution 
[Meakin, Feder and J\o ssang \cite{f}],
\[
P_{\partial}(a,L) = a^{- \tau_{\partial}} f_{\partial} 
\biggl( \frac{a}{L^{1+H}} \biggr)
\]
then, as in (\ref{amedia}), one finds:
\[
\langle a \rangle_{\partial} \sim L^{(1+H)(2-\tau_{\partial})} \, .
\]
As noted in \cite{f} the mean accumulated area for a border 
site in a basin is just the number of sites in the bulk divided by the number 
of sites in the border. Denoting by $D$ the fractal dimension of the border,
\[
\langle a \rangle_{\partial} L^{D} \sim L^{(1+H)} \, ,
\]
$\tau_{\partial} = 1+ \frac{D}{1+H} $. In the case of compact basins $(H=1)$,
$\tau_{\partial} = 1+ \frac{D}{2}$, ( $ \tau_{\partial} = 3/2 $ for a square ).
Note that $\tau_{\partial}$ is independent of the class of spanning tree
chosen to drain the basin and it is $ a \,\, priori$ different from the $\tau$
we introduced to characterize the distribution of accumulated areas in the 
bulk, which we found to be for $H=1$, $\tau = 2 - \frac{d_{l}}{2}$.

\bibitem{an}
D. A. Huse and C. L. Henley, {\it Phys. Rev. Lett.} 
{\bf 54}, 7708 (1985).

\bibitem{ao}
 M. Kardar, {\it Phys. Rev. Lett.} 
{\bf 55}, 2923 (1985).

\bibitem{ap}
D. A. Huse, C. L. Henley and D. S. Fisher, {\it Phys. Rev. 
Lett.} {\bf 55}, 2924 (1985).

\bibitem{at}
A. Rinaldo, A. Maritan, F. Colaiori, A. Flammini, M. Swift, 
 R. Rigon, J. R. Banavar and I. Rodriguez-Iturbe,
submitted to {\it Nature}.

%\bibitem{e} 
%L. B. Leopold and W. B. Langbein, U. S. Geol. Surv. Prof. Pap. 
%{\bf 500-A} (1962).

%\bibitem{l} 
%I. Rodriguez-Iturbe, R. L. Bras, E. Ijjasz-Vasquez, D. G. Tarboton, 
%{\it Water Resour. Res.} {\bf 28}, 988 (1992).

%\bibitem{o}
%S. P. Breyer and R. S. Snow, 
%Geomorphology {\bf 5}, 143 (1992).

%\bibitem{q}
%D. R. Montgomery and W. E. Dietrich, 
%{\it Science} {\bf 255}, 826 (1992).

%\bibitem{s}
%R. E. Horton, Geol. Soc. America Bull. 
%{\bf 56}, 275 (1945).

%\bibitem{t}
%R. E. Horton, Eos Trans. AGU 
%{\bf 13}, 350 (1932).

%\bibitem{u}
%R. L. Shreve, J. Geol. 
%{\bf 74}, 17 (1966).

%\bibitem{v}
%R. L. Shreve, J. Geol. 
%{\bf 75}, 178 (1967).

%\bibitem{z}
%R. L. Shreve, J. Geol. 
%{\bf 77}, 397 (1969).

%\bibitem{w}
%A. D. Howard, Water Res. Res. 
%{\bf 26}, 2107 (1990).


%\bibitem{y}
%A. D. Howard, Water Res. Res. 
%{\bf 7}, 863 (1971).

%\bibitem{aa}
%C. T. Yang, Water Res. Res. 
%{\bf 7}, 311 (1971).

%\bibitem{ab}
%P. S. Stevens, "Patterns in Nature", 
%Little-Brown, Boston Mass. (1974).



\end{thebibliography}
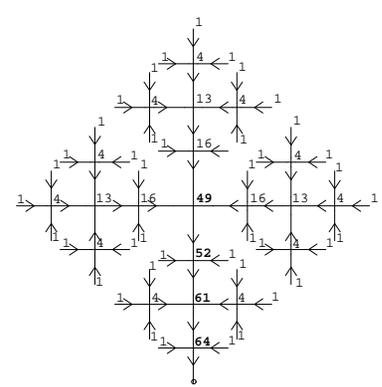
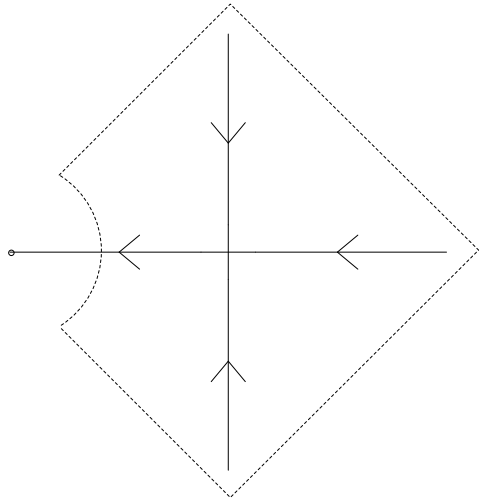
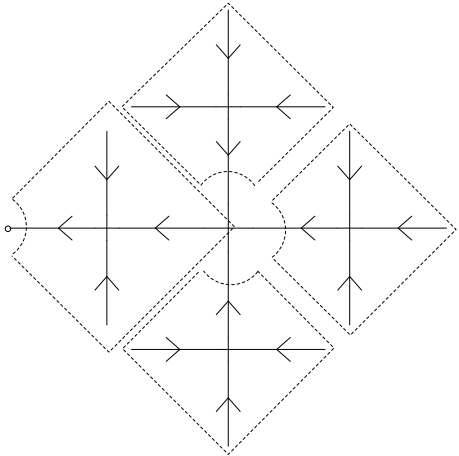
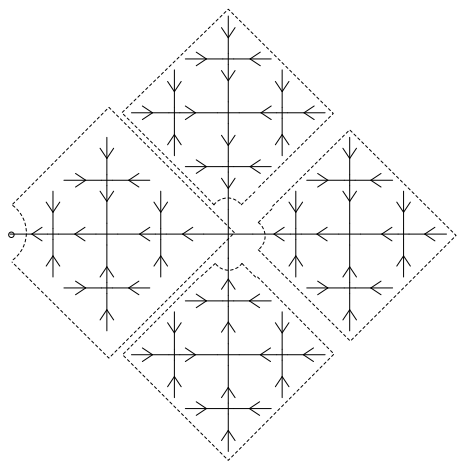
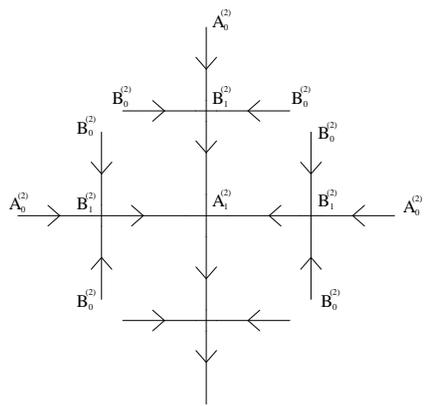
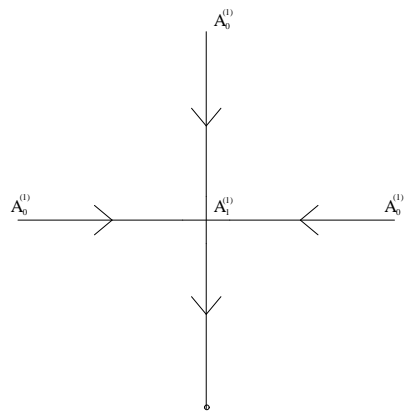
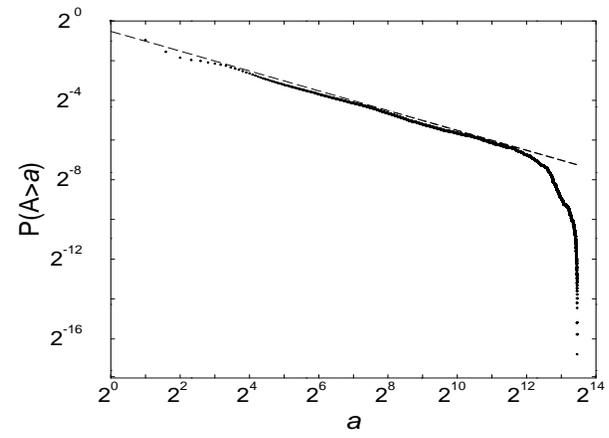
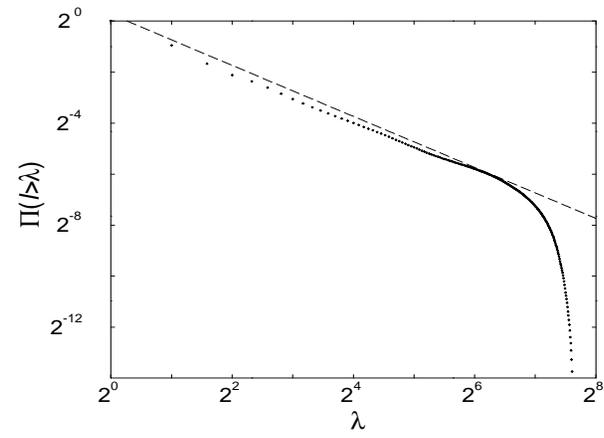
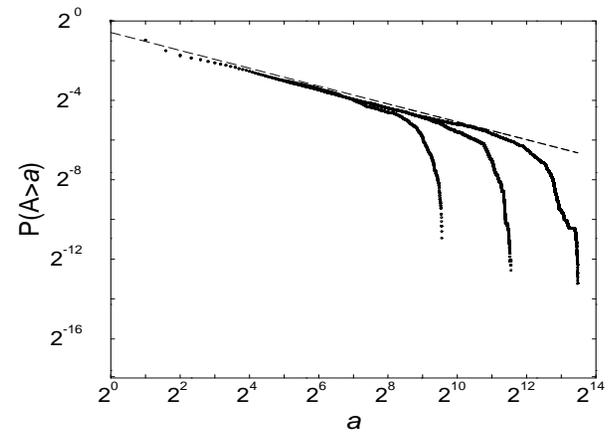
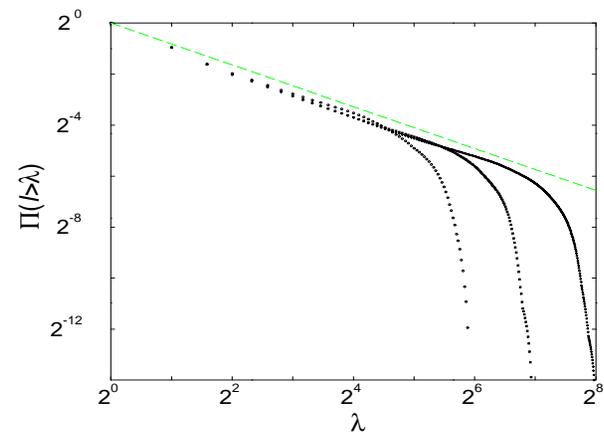
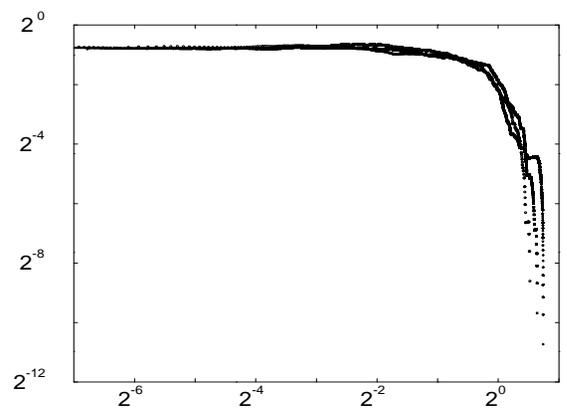
\end{document}